%% file: main.tex
\documentclass[handcarry]{aiaa-tc}

\graphicspath{{./figures/}}

\title{GTOC8:  Results and Methods of ESA Advanced Concepts Team and JAXA-ISAS}

\author{Dario Izzo, Daniel Hennes, Marcus M\"artens \\ Ingmar Getzner, Krzysztof Nowak, Anna Heffernan \\ Stefano Campagnola, Chit Hong Yam, \\ Naoya Ozaki, Yoshihide Sugimoto}

 \author{
  Dario Izzo\thanks{Scientific Coordinator, Advanced Concepts Team, ESA},
  \ Daniel Hennes\thanks{Research Fellow, Advanced Concepts Team, ESA.}, \ Marcus M\"artens\thanks{Ph.D. Candidate, TU Delft.}, \\ Ingmar Getzner\thanks{Young Graduate Trainee, Advanced Concepts Team, ESA.}, \ Krzysztof Nowak\thanksibid{4}, \ Anna Heffernan\thanksibid{2}\\
  {\normalsize\itshape
   ESA's Advanced Concepts Team, Noordwijk, 2201 AZ, The Netherlands}\\
  \and
  Stefano Campagnola\thanks{International Top Young Fellow, Department of Space Flight Systems, JAXA.},
   \ Chit Hong Yam\thanks{JSPS Postdoctoral Research Fellow, Department of Space Flight Systems, JAXA.}, \\ Naoya Ozaki\thanks{Ph.D. Candidate, The University of Tokyo}, Yoshihide Sugimoto\thanksibid{6} \\
  {\normalsize\itshape
  ISAS/JAXA, Yoshinodai 3-1-1,Sagamihara, Kanazawa,  Japan}
 }

 \AIAApapernumber{16-275}
 \AIAAconference{AAS/AIAA Space Flight Mechanics Meeting, Napa, CA in February 14-18, 2016}

 \newcommand{\argmin}{\operatornamewithlimits{argmin}}
 \newcommand{\argmax}{\operatornamewithlimits{argmax}}
 \newcommand{\degg}{[deg] }

\usepackage[english]{babel}
\usepackage[utf8]{inputenc}
\usepackage{amsmath}
\usepackage{graphicx}
\usepackage[noend]{algpseudocode}
\usepackage{xcolor}
\usepackage[colorinlistoftodos]{todonotes}
\usepackage{wrapfig}
\usepackage{algorithm}
\usepackage{enumitem}
\usepackage{tikz}
\usepackage{multirow}

\usetikzlibrary{3d,calc}

\begin{document}

\maketitle

\begin{abstract}
We consider the interplanetary trajectory design problem posed by the 8th edition of the Global Trajectory Optimization Competition and present the end-to-end strategy developed by the team ACT-ISAS (a collaboration between the European Space Agency's Advanced Concepts Team and JAXA's Institute of Space and Astronautical Science). The resulting interplanetary trajectory won 1st place in the competition, achieving a final mission value of $J=146.33$ [Mkm]. Several new algorithms were developed in this context but have an interest that go beyond the particular problem considered, thus, they are discussed in some detail. These include the Moon-targeting technique, allowing one to target a Moon encounter from a low Earth orbit; the 1-$k$ and 2-$k$ fly-by targeting techniques, enabling one to design resonant fly-bys while ensuring a targeted future formation plane
; the distributed low-thrust targeting technique, admitting one to control the spacecraft formation plane at 1,000,000 [km]; and the low-thrust optimization technique, permitting one to enforce the formation plane's orientations as path constraints.
\end{abstract}

\section*{Nomenclature}

\begin{tabbing}
  XXXXXXXXXXX \= \kill
  $a$ \> Orbit semi-major axis \\
  $e$ \> Orbit eccentricity \\
  $J$ \> Mission value (performance index) \\
  $\delta$ \> Declination of a radio source\\
  $\Delta t_0$ \> Starting time of \emph{Phase 3b}\\
  $\Delta t_w$ \> Upper time bound on the \emph{Phase 3b} search for potential targets along celestial track\\
  $\Delta t_i^{-}$ \> Approximate observation time in \emph{Phase 3b} \\
  $\Delta t_i^{*}$ \> Observation time in \emph{Phase 3b} \\
  $\mathcal O$ \> List of radio-sources observed during \emph{Phase 3b}\\
  $V$ \> Previously observed radio sources of \emph{Phase 3b}\\
  $V^+$ \> Solutions of \emph{Phase 3b}\\
  $h$ \> Smallest of the three altitudes of the observing triangle \\
  $\mu$ \> Moon gravitational parameter \\
  $P$ \> Multiplicative factor \\
  $t_0$ \> Mission start, start of a trajectory Phase\\
  $\mathbf r$ \> Position vector \\
  $r$ \> Position vector magnitude\\
  $r_m$ \> Closest Moon safe distance\\
  $N$ \> Total number of observations\\
  $\mathbf n$ \> Radio-source direction\\
  $n$ \> Potential radio-source in \emph{Phase 3b}\\
  $h_\mathcal{O}$ \> Baselines of a radio-sources observation \emph{Phase 3b}\\
  $\mathcal S_1, \mathcal S_2, \mathcal S_3$ \> The three spacecraft, or their Keplerian orbits.\\
  $\mathcal S$ \> Shorthand for all three spacecraft $\mathcal S_1, \mathcal S_2, \mathcal S_3$\\
  $\mathbf v$ \> Velocity vector \\
  $V$ \> Velocity vector magnitude \\
  $P_1$, $P_2$, $P_3$ \> Virtual planets \\
  $RA$ \> Right ascension of a radio source\\
  $\mathbf T$ \> Thrust vector\\
  $\mathcal T$ \> List of resonances considered\\
  $\tau$ \> Orbital period\\
  $\mathbf u$ \> Thrust vector direction\\
  $\overline T$ \> Thrust in the outgoing spirals\\
  $T_{max}$ \> Maximum allowed thrust\\
  $\mathcal P$ \> Keplerian celestial track\\
  $\theta_1, \theta_2$ \> Anomalies used in the design of the spirals\\
  $i$ \> Inclination\\
  $\hat{\boldsymbol i}_v$, $\hat{\boldsymbol i}_h$, $\hat{\boldsymbol i}_n$ \> Orbital frame \\
  $\alpha$ \> Pull angle\\
  $k$ \> Crank angle\\ [5pt]
  
  \textit{Subscripts}\\
  $\infty$ \> Velocity relative to the Moon \\
  $SC$ \> Refers to the spacecraft \\
  $M$ \> Refers to the Moon \\
  $AN$ \> Refers to the ascending node \\
  $DN$ \> Refers to the descending node \\ [5pt]
  
  \textit{Superscripts}\\
  $-$ \> Before a Moon encounter \\
  $+$ \> After a Moon encounter \\

 \end{tabbing}

\section{Introduction}
The 8th edition of the Global Trajectory Optimization Competition (GTOC8) was organized by Jet Propulsion Laboratory in the first half of 2015. In this paper, the end-to-end strategy developed by the team ACT-ISAS (a collaboration between the European Space Agency's Advanced Concepts Team and JAXA's Institute of Space and Astronautical Science) is described. Some of the authors had previously developed complex interplanetary trajectories in the framework of other GTOC competitions \cite{gtoc5, gtoc6} as members of the same team.
The problem considered by GTOC8 was that of a \lq\lq high-resolution mapping of radio sources in the universe using space-based Very-Long-Baseline Interferometry (VLBI)\rq\rq\;. Its mathematical formalization can be read in the work written by the competition organizer, Anastassios Petropoulos\cite{gtoc8description}. We refer the reader to this paper for all formal details on the interplanetary trajectory problem studied in detail here. 

\section{Preliminary Analysis}
\input{idealgeometry}
\noindent
An overall design strategy, for the GTOC8 problem, is designed having in mind the maximization of the mission value defined as \cite{gtoc8description}:
$$
J = \sum_{j=1}^N P_ih_i(0.2 + \cos^2\delta_i), 
$$
where $h_i \ge 10000$ [km] is the observation baseline, defined as the smallest of the three altitudes of the observing triangle (the triangular formation acquired by the three satellites at the time of the $i$-th observation); $P_i \in \left\{1,3,6\right\}$ is the multiplicative factor that favours increasing $h$ values when repeating observations; and $\delta_i$ is the latitude of the observed radio source. While the exact values of these numbers at each observation, and thus of $J$, will be the result of a complex trajectory design, it is useful to note early-on how low latitude observations are highly favoured as well as large values of $h$ connected with a large $P=6$ multiplier (note that this implies that the same radio source is observed twice before with increasingly smaller baselines\cite{gtoc8description}).

\subsection{The ideal formation}
\label{sec:ideal}
Given the constraint $r < 1,000,000$ [km] on the maximum magnitude of the spacecraft position vector, it follows that the maximum value for the baseline, $h$, of an observation (corresponding to an equilateral triangle with vertexes on the allowed domain border) is 1,500,000 [km]; hence, the maximum value of one single ideal observation (taken with $P=6$) is $10,800,000$ [km]. For 
the occurrence of the ideal case, the same radio source (having $\delta = 0$) must have been scored twice already, delivering at most a value of $200,000$ [km] at its first observation and $1,800,000$ [km] at its second. In general, for any given $P=6$ observation, a maximum of only $\frac{5}{27} \approx 20\% $ of its value can be delivered by the previous two observations. This suggests to implement a solution strategy where the number of $P=6$ observations is maximized, while the $P=1$ and $P=3$ observations are only used to enable $P=6$ observations, and not, at least in a first instance, to build up the overall mission value. The ideal observation discussed above can be obtained repeatedly placing the three satellites on circular orbits with semi-major axis equal to $1,000,000$ [km]. Two options, shown in Figure \ref{fig:ideal}, are possible: three orbits at 90 \degg of inclination and three orbits at 0, 60 and 60 \degg inclination, the latter two orbits having a RAAN difference of 180 \degg. In the first case, the ideal formation is kept during the whole orbit, while in the second case it is only acquired twice per orbit, corresponding to, roughly, once every 57.6 days. As we do not want to repeat the same observation more than once (under the hypothesis that the $P=1$ and $P=3$ observations are already done), we will have to modify the ideal geometry using thrust to allow the targeting of different radio sources. We observe that this requires a larger thrust effort in the first case, where the formation plane has no natural dynamics and is fixed. Also, any re-targeting manoeuvre in the first case requires orbital inclination changes, while in the second case the formation plane's orientation targeting can be achieved 
by merely changing the phase of $\mathcal S_2, \mathcal S_3$ (see Figure \ref{fig:ideal}). The 0, 60 and 60 \degg is then considered as the ideal orbital geometry.

\subsection{The chosen design strategy}
\label{sec:baseline}
The design strategy chosen divides the overall mission in 3 main phases and is aimed at maximising the number of $P=6$ observations with $h$ as close to its ideal value as possible. Alternative designs are feasible and briefly discussed in a later section of the paper.

\subsubsection{Phase 1 - Get to the Moon}
The three spacecraft start their journey on an equatorial circular orbit at an altitude of 400 [km]. The initial $\Delta V$ available from the launcher, a maximum 3 [km/s], does not suffice to send them directly to a Moon encounter. Hence, a spiralling out phase is required, which we refer to as \emph{Phase 1} or, simply, ``Get to the Moon''. This first phase starts at $t_0$ and ends after $T_1$ when the first of the three spacecraft arrives at a Moon encounter. The Moon encounter relative velocities eventually achieved are indicated with $V_\infty^{\mathcal S_1}$, $V_\infty^{\mathcal S_2}$, and $V_\infty^{\mathcal S_3}$. As our goal is to maximise the number of $P=6$ observations, it follows that phase 1 has to be completed as quickly as possible, resulting in a time-optimal problem. No observations are made during this phase negating a requirement for out of plane motion, thus, the Moon encounters occur at the Moon ascending and descending nodes, which in turn, allows for a simplified spiralling out design.

\subsubsection{Phase 2 - Play the Moon}
After the first spacecraft arrives at the Moon, the second trajectory phase starts, which we refer to as \emph{Phase 2} or, simply, ``Play the Moon''. At each Moon encounter, each spacecraft performs a Moon fly-by designed to put the spacecraft in a Moon resonant orbit. This guarantees a further Moon encounter while also targeting a radio-source observation at some successive epoch. We will refer to this fly-by as a $k$-targeting manoeuvre, after the variable $k$ used to denote the crank angles chosen to target an observation, as detailed later. Note that depending on the details of the trajectory, at each node there could be one, two or three spacecraft performing a fly-by. We refer to these cases as 1-$k$, 2-$k$ and 3-$k$ targeting manoeuvres, but we will only use 1-$k$ and 2-$k$ targeting as it is convenient to occupy both Moon nodes. Assuming that at each successive Moon ascending and descending node, there is at least one spacecraft performing a fly-by, there will be a $k$ targeting manoeuvre approximately each $13.6$ days.  This seems like a good frequency given the constraint of $15$ days between successive observations. 

\subsubsection{Phase 3a - Get out}
Having observed a sufficient amount of radio sources with multiplicative factors $P=1$ and $P=3$, the spacecraft use one last Moon fly-by to start a transfer that targets the ideal orbital geometry of 0, 60 and 60 \degg inclination.  We refer to this as \emph{Phase 3a} or, simply, ``Get out''. During \emph{Phase 2}, no propulsion is used, thus the mass of the spacecraft is unchanged, as well as their relative Moon encounter velocities $V_\infty^{\mathcal S_1}$, $V_\infty^{\mathcal S_2}$, and $V_\infty^{\mathcal S_3}$. The Moon velocity is $V_{M_{AN}} = 1039.77$ [km/s] at the ascending node and $V_{M_{DN}} = 1003.98$ [km/s] at the descending node. As a consequence, the maximum inclination that the spacecraft can acquire after the last Moon fly-by of \emph{Phase 2} is:
\begin{equation}
\label{eq:maxincl}
i = i_{M} \pm \arcsin\left(\frac{V_\infty}{V_M}\right),
\end{equation}
where the $\pm$ corresponds to the two possible extremal exit conditions (up and down). As an indication, the arcsin contribution ranges from 30 to 64 \degg as $V_\infty$ increases from $550$ to $900$ [m/s]. The spacecraft propulsion system during \emph{Phase 3a} is, thus, mainly used to circularize the orbit at 1,000,000 [km] and to acquire the correct ideal phasing; the inclination change can be taken care of by the initial fly-by if we assume two spacecraft have enough $V_\infty$. During \emph{Phase 3a}, some further radio sources can be observed, but only a few options are available as the main trajectory goal is to acquire the target ideal geometry in the minimum time.

\subsubsection{Phase 3b - The million kilometer game}
Once the three spacecraft have been inserted in their final ideal orbital configuration, the low-thrust propulsion system will be used to control the formation plane.  They, thus, observe again the radio sources previously observed (mainly) during \emph{Phase 2} but now with a baseline large enough to allow for the multiplicative factor $P=6$. This is what we refer to as \emph{Phase 3b} or, simply, ``The million kilometer game''.
\vskip0.5cm
\noindent
The described design strategy relies on the fact that the same radio sources are observed at low and medium baselines during \emph{Phase 2} and \emph{Phase 3a}, and then again during \emph{Phase 3b}, but now with a large enough baseline so as to allow for the multiplicative factor $P=6$. A simple \lq\lq trick\rq\rq\ ensures this happens without introducing any further complexity: having designed \emph{Phase 1} and, thus, having computed $V_\infty^{\mathcal S_1}$, $V_\infty^{\mathcal S_2}$, $V_\infty^{\mathcal S_3}$ (as during \emph{Phase 2} these and the spacecraft masses will remain unchanged), we may start designing \emph{Phase 3} with the aim to observe as many different radio sources as possible with large baselines. We may then design \emph{Phase 2} as a last step, trying to target the same radio observations with smaller baselines.

\section{Phase 1 - Get to the Moon}
\input{spirals}
During the first phase, the three spacecraft travel to encounter the Moon at one of the nodes using a planar transfer. In order to design such a transfer, 
we assume the thrust control law in the form $\mathbf T = \overline T \mathbf u$, where $\mathbf u$ is a unit vector and $\overline T \le T_{max}$ a constant parameter. The spacecraft dynamics (two dimensional) is thus:
\begin{equation}
\begin{array}{l}
    \ddot{\mathbf r} = -\frac{\mu}{r^3}\mathbf r + \frac{\overline T}{m}\mathbf u, \\
    \dot m = -\frac{\overline T}{I_{sp}g_0}.
\end{array}
\end{equation}
Assuming a starting position on the x-axis, say $\mathbf r_0 = [-r_0, 0]$, with velocity, $\mathbf v_0 = [0, -v_0]$, we numerically propagate the above equations, halting the numerical integration when either $r = r_{AN}$ or $r = r_{DN}$, according to the targeted Moon node passage (without loss of generality we assume the ascending node is chosen here).  
While $r_0$ is the parking orbit radius, we compute $v_0$ as the circular velocity on the parking orbit plus the launcher $\Delta V = 3000$ [m/s]. The anomaly $\theta_2$, defined in Figure \ref{fig:spiral}, is then easily computed as $\theta_2 = \pi + \arctan 2(y_f, x_f)$, where $x_f, y_f$ are the Cartesian coordinates of the final condition reached, taken to be on the line of nodes. From the resulting time of flight, $t_f$, we also compute $\theta_1$ as the anomaly between the line of nodes and the spacecraft position vector along its initial parking orbit at $t_{\mbox{AN}} - t_f$, where $t_{\mbox{AN}}$ is the epoch of the targeted Moon ascending node passage. As shown in Figure \ref{fig:spiral}, by construction, if $\theta_1 + \theta_2 = 2\pi$ the spiral transfer is feasible. This suggests to design the outgoing spirals by choosing a thrust law parametrized by $\overline T$ and finding the value of $\overline T \le T_{max}$ with a zero finding method so that,
\begin{equation}
\theta_1(\overline T) + \theta_2(\overline T) = 2\pi.
\label{eq:condition}
\end{equation}
Three control laws defining the thrust direction $\mathbf u$ are considered.
\subsubsection*{a. Tangential control law}
The spacecraft is constantly keeping a perfectly tangential thrust so that:
\begin{equation}
       \mathbf u = \hat{\boldsymbol i}_v,
\end{equation}
where $\hat{\boldsymbol i}_v$ is a unit vector aligned with the spacecraft velocity. Under this control law, once we fix the Moon node passage targeted, Eq.(\ref{eq:condition}) admits infinitely many solutions; each one corresponding to a different number of revolutions needed to first reach the node distance. In our case, we are mainly interested in getting to the moon in the shortest possible time, so the minimum number of revolutions can be selected, and a unique spiral is thus computed for each target Moon node passage.

\subsubsection*{b. Fixed perigee control law}
The tangential control law results in a transfer that encounters the moon at a low relative velocity (on the order of 550 [m/s]), as the transfer orbit perigee is also increased during the transfer. In order to find transfers that allow for larger relative velocities at the Moon encounter and, thus, higher inclinations after a Moon fly-by, see Eq.(\ref{eq:maxincl}), we consider a second control law that is guaranteed to keep the orbit perigee fixed while increasing its apogee\cite{campagnola2014}. Such a control law has the form:
\begin{equation}
    \left\{ \begin{array}{ll}
            \mathbf u = [1,0] & f = 0 \\
            \mathbf u = \mbox{sign}(\sin f) \frac{\mathbf p}{|p|} & f \ne 0
    \end{array} \right.
\end{equation}
where $\mathbf p = r \sin f \hat{\boldsymbol i}_v + [2a(\cos f+1) -4a^2(1-e)/r] \hat{\boldsymbol i}_n$. As above, under this control law infinite solutions exist, and we select the one having the minimum number of revolutions (minimum time); hence, we obtain a unique spiral for each target Moon node passage. The resulting relative encounter speed is increased to the order of 900 [m/s].

\begin{figure}[t!]
\centering
\includegraphics[width=0.32\columnwidth]{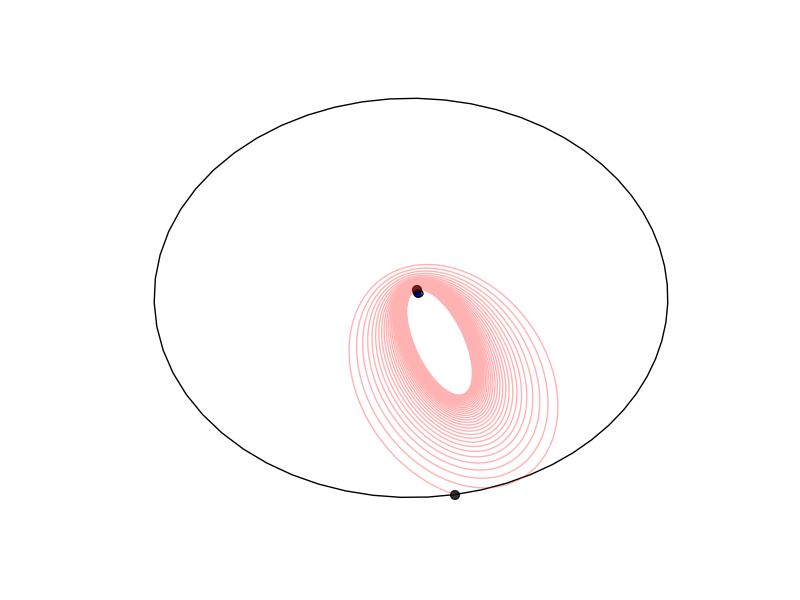}
\includegraphics[width=0.32\columnwidth]{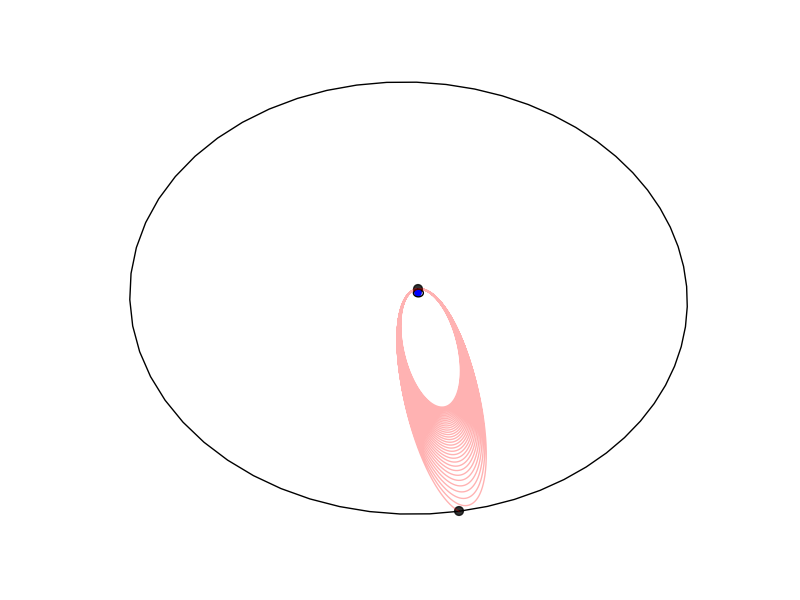}
\includegraphics[width=0.32\columnwidth]{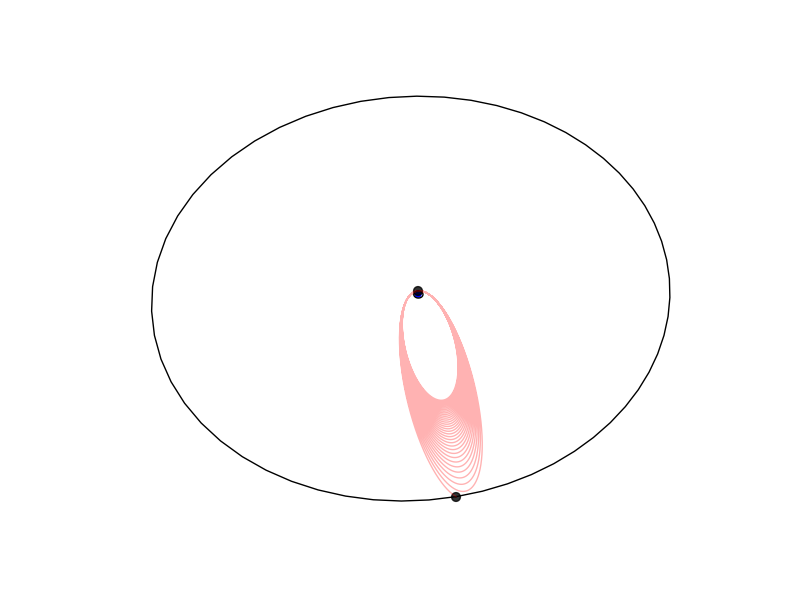}
\caption{Different spirals reaching the Moon at the descending node. Purely tangential thrust (left), thrust inversion at $\nu=3.75$ \degg (center) and fixed perigee control law (right). Encounter $v_{\infty}$ are, for these spirals, $555$, $824$ and $829$ [m/s]. \label{fig:spirals}}
\end{figure}

\subsubsection*{c. Inversion control law}
In order to allow for greater flexibility in the selection of the final Moon encounter relative velocity, we consider a third control law that keeps the thrust as tangential but inverts its direction within an angle $\nu$ from the orbit apogee. Formally:
\begin{equation}
    \left\{ \begin{array}{ll}
            \mathbf u =  \hat{\boldsymbol i}_v & f\in[\nu-\pi, \pi-\nu] \\
            \mathbf u = -\hat{\boldsymbol i}_v & f \notin [\nu-\pi, \pi-\nu]
    \end{array} \right.
\end{equation}
Such a control law, for a fixed $\nu$, also allows the computation of a unique spiral for each target Moon node passage; however, by adjusting $\nu$, we may vary continuously the final encounter relative velocity gaining some design flexibility. 

\vskip0.5cm
\noindent
For illustration purposes, we visualize some spiralling out trajectories in Figure \ref{fig:spirals}, which are obtained for the same descending node Moon passage using the different design techniques. We note that the same technique could be applied with  control laws defined using more than one parameter at the cost of increasing the complexity of numerically finding the solution to the equivalent of Eq.(\ref{eq:condition}).

\section{Phase 3 - 1,000,000 km}
The third phase of the trajectory is where the baseline of the interferometric measurements is the highest. Its design only requires $V_\infty^{\mathcal S_1}$, $V_\infty^{\mathcal S_2}$, $V_\infty^{\mathcal S_3}$ and the information about the node they refer to, e.g. [AN(+1), AN(+1), DN] indicates that $\mathcal S_3$ performs its last fly-by at the descending node and $\mathcal S_1, \mathcal S_2$ at the following ascending node. The exact epoch of the node Moon passages is decided later when the whole trajectory is assembled. The whole of \emph{Phase 3} is, thus, designed in a relative timeline. This is possible as radio-sources are fixed in the celestial sphere; hence, we will be able to move the start of \emph{Phase 3} to any node passage epoch as long as we respect the node sequence (i.e. [AN(+1), AN(+1), DN]).
The idea is to first acquire the ideal orbital geometry, visualized in Figure \ref{fig:ideal}, consisting of three satellites in circular orbits at inclination 0\degg, 60\degg and 60\degg; thus, they form an equilateral triangle of maximal side twice per orbit (\emph{Phase 3a}). Once we have reached this final target configuration, the on-board propulsion is used to target as many different radio-source observations as possible 
in the shortest time (\emph{Phase 3b}).

\subsection{Phase 3a - Get out}
\begin{figure}[t!]
\centering
\includegraphics[width=0.4\columnwidth]{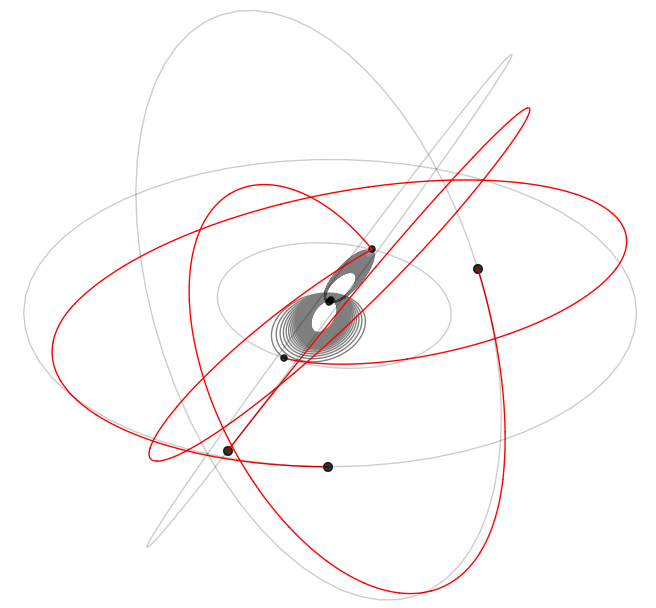}
\includegraphics[width=0.59\columnwidth]{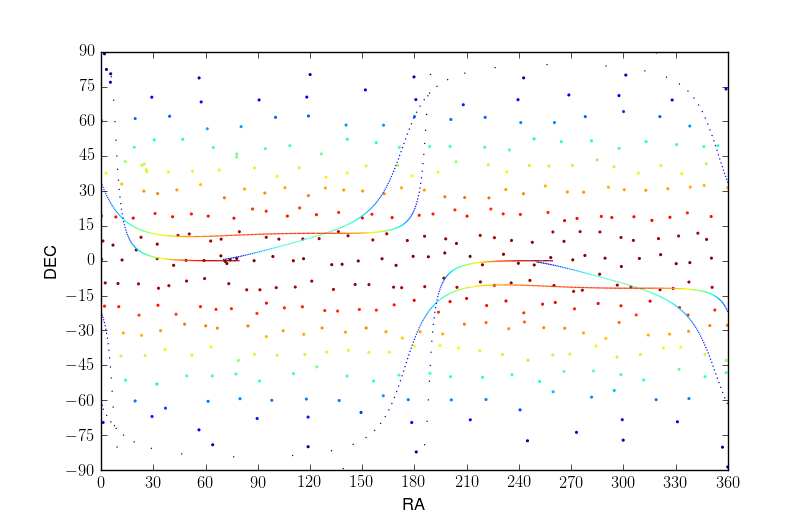}
\caption{Example of the outcome of \emph{Phase 3a} design. The outgoing trajectories (left) and their celestial track (right) before any radio-source observation is enforced. \label{fig:phase3a}}
\end{figure}
In order to acquire the 
ideal orbital configuration, three spacecraft transfers are designed in sequence, using as initial conditions 
the output from the three spiral transfers 
in \emph{Phase 1}. At this stage, \emph{Phase 2} is neglected as the three Moon encounter velocities, $V_\infty^{\mathcal S_i}$, and spacecraft masses, $m_i$, as well as the nodes where the encounters happen, are invariant. The low-thrust, time-optimal control problem of a spacecraft performing a Moon fly-by at $t_0$ while targeting a final \lq\lq virtual\rq\rq\ planet (subscript $P$) is considered and approached using a direct optimization method\cite{PyKEP}. The whole transfer trajectory is divided into $n_{seg}$ segments of equal duration, along which the spacecraft thrust is assumed to be inertially constant.  This allows us to define  the following non linear programming problem (NLP),

\begin{equation}
    \begin{array}{rl}
         \mbox{find:} & [k, \alpha, tof, t_{P}, m_f, \mathbf u_1, .. \mathbf u_{n_{seg}}]  \\
         \mbox{to minimize:} & tof
         \\
         \mbox{subject to:} & \mathbf r(t_0+tof) = \mathbf r_P(t_P)
         \\
                            & \mathbf v(t_0+tof) = \mathbf v_P(t_P)
         \\
                            & \mathbf u_i \in [0,1], \quad \forall i = 1 .. n_{seg}
    \end{array},
\end{equation}
where $tof$ is the transfer time-of-flight; $m_f$ is the final spacecraft mass; $\mathbf u_{i}$ is the \lq\lq throttle\rq\rq\ defining the spacecraft's thrust at segment $i$ as $T_i=T_{max} \mathbf u_i$; and $t_P$ is the virtual planets ($P_1$) reference epoch, essentially dictating the position of the virtual planet along its orbit. We first solve the above problem starting from the $\mathcal S_1$ conditions at the end of \emph{Phase 1} and defining the target planet to be on a 60\degg inclined circular orbit with a semi-major axis $a = 990000$ [km]. We then consider a modified form of the NLP:

\begin{equation}
    \begin{array}{rl}
         \mbox{find:} & [k, \alpha, tof, m_f, \mathbf u_1, .. \mathbf u_{n_{seg}}]  \\
         \mbox{to minimize:} & tof
         \\
         \mbox{subject to:} & \mathbf r(t_0+tof) = \mathbf r_P(t_0+tof)
         \\
                            & \mathbf v(t_0+tof) = \mathbf v_P(t_0+tof)
         \\
                            & \mathbf u_i \in [0,1], \quad \forall i = 1 .. n_{seg}
    \end{array},
    \label{eq:NLP2}
\end{equation}
where the virtual planet reference epoch, $t_P$, no longer appears as we aim at a precise point along the target orbit. The above problem is solved starting from the $\mathcal S_2$ conditions at the end of \emph{Phase 1} and considering a virtual planet $P_2$, defined by:
$$
\begin{array}{l}
     \mathbf r_{P_2}(t) = [x(t), y(t), -z(t)]  \\
     \mathbf v_{P_2}(t) = [v_{x_f}(t), v_{y_f}(t), -v_{z_f}(t)] 
\end{array},
$$
where $t$ is any epoch, while $x,y,z, v_x, v_y, v_z$ are the Cartesian components of $\mathcal S_1$ position and velocity, computed at $t$ along its target 60\degg inclined circular orbit. Note that $P_2$ will thus be on a 60\degg inclined circular orbit at 990000 [km] that has an offset of 180\degg in the right ascension of the ascending node. To design the third transfer, concerning the spacecraft $\mathcal S_3$, we solve again the NLP in Eq.(\ref{eq:NLP2}) but starting, this time, from the $\mathcal S_3$ conditions at the end of \emph{Phase 1} and targeting a virtual planet $P_3$. $P_3$ has a circular orbit of 0\degg inclination, a semi-major axis of $a = 990000$ [km], and a phase such that a maximal triangle is achieved twice per orbit when considering the formation made by the the three virtual planets $P_1$, $P_2$ and $P_3$. The outcome of these three low-thrust optimizations is visualized, for one particular case, in Figure \ref{fig:phase3a}. The celestial track, that is the projection on the celestial sphere of the unit vector $\mathbf n$ normal to the $\mathcal S_1$, $\mathcal S_2$, $\mathcal S_3$ formation plane, is also shown. The celestial track shows clearly how the three satellites, during this phase, may come very close to make a number of radio-observations. To enforce a few of them (selected manually), the trajectory of $\mathcal S_2$ is re-optimized adding one path constraint for each forced observation: $\left(\mathbf r_3(t^*) - \mathbf r_2(t^*)\right) \cdot \mathbf n = 0$, where $t^*$ is the observation epoch computed previously solving $\left(\mathbf r_1(t^*) - \mathbf r_3(t^*)\right) \cdot \mathbf n = 0$ for $t^*$.

\subsection{Phase 3b - The million kilometer game}

When the last of the three spacecraft acquires its target final orbit at $a = 990000$ [km], a new phase begins. Now, the on-board low-thrust propulsion is used to target different large baseline radio observations so as to maximise $J$ (assuming $P=1$) in a compact schedule. The multiplicative factors will later be increased (e.g. to $P=6$) via the design of \emph{Phase 2}. At the end of \emph{Phase 3b}, we compute a set of radio-sources, $\mathcal O$; baselines, $h_\mathcal O$; and the times, $\Delta t^*_i$, at which such radio-sources are observed after the start of this phase. These times will be converted to epochs when the whole trajectory is assembled and the exact ascending and descending node Moon passage epochs will be known for the \emph{Phase 3a} beginning.

\subsubsection{Distributed low-thrust targeting technique}
Consider $\mathbf r_{0_j}, \mathbf v_{0_j}, m_{0_j}$ as the state of $\mathcal S_j$ at a given $\Delta t^*_i$. In order to design a low-thrust transfer that will lead to a further observation, we consider the thrust of the three spacecraft as fixed in our inertial frame and we denote it with $\mathbf u_j$. The following NLP is then considered:
\begin{equation}
    \begin{array}{rl}
          \mbox{find:} & \mathbf u_1, \mathbf u_2, \mathbf u_3, T^* \\
          \mbox{to minimize:} & J = \sum_i |\mathbf u_i|\\
          \mbox{subject to:} & |\boldsymbol\eta(T^*) \cdot \mathbf n| \le \epsilon \\
                             & r_1, r_2, r_3 < 1000000 [km] 
    \end{array},
\end{equation}
where $\boldsymbol\eta(T^*) = \frac{\mathbf r_{12}(T^*) \times \mathbf r_{23}(T^*)}{|\mathbf r_{12}(T^*) \times \mathbf r_{23}(T^*)|}$ is the formation plane's orientation after the three spacecraft have thrusted for a period $T^*$ using the three fixed values $\mathbf u_1, \mathbf u_2, \mathbf u_3$. The objective function, $J$, represents some sort of distributed minimum mass consumption. Alternatively, a maximum baseline $h$ could be targeted, but this leads to the likely violation of the $r < 1000000$ [km] constraint and to a fast departure from the ideal orbital geometry. The spacecraft state propagation is efficiently done using a Taylor integration scheme \cite{taylorint, YamICATT, PyKEP}, where the path constraint on $r_i < 1000000$ [km] is not enforced but only checked after the optimization completes. When violated, that particular radio-source observation is not included in the list of options for a further observation. The resulting NLP is then rather simple and can be solved for multiple candidate radio-source directions $\mathbf n$ as well as inserted in a more complex tree search as described in the following section.

\begin{figure}[t!]
\centering
\includegraphics[width=0.79\columnwidth]{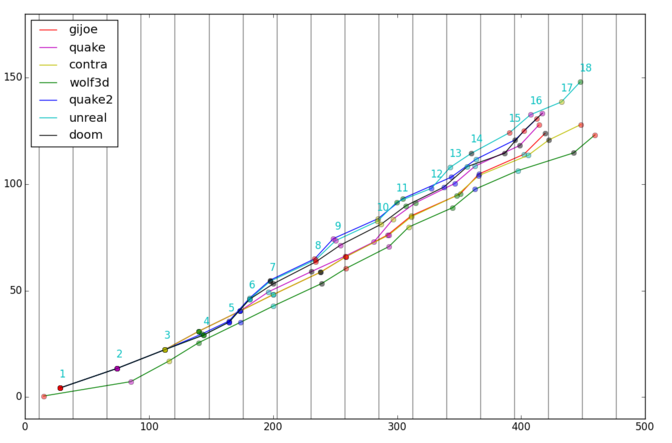}
\caption{Comparison of different results from runs of the low-thrust targeting tree search (each partial trajectory is named after a famous video-game shown in the legend). The cumulative mission value in [km] is reported on the $y$-axis ($P=1$ assumed), while the trajectory duration in [days] is shown on the $x$ axis. The vertical lines correspond to the Moon node passage (ascending). Radio-source observations are represented as coloured circles. \label{fig:lt_target}}
\end{figure}

\subsubsection{The low-thrust targeting tree search}
We use a DFS-inspired tree-search to optimize the targeting maneuvers in this sub-phase. Each node of the tree represents a radio source observed using low-thrust targeting. The general evaluation and branching loop can be described in two steps:

\begin{enumerate}
    \item Retrieve the information on the search state (spacecraft states, time, observed targets thus far and next potential observation).
    \item Evaluate the search state and branch into further states (target the observation, propagate the spacecraft trajectories and scan the celestial-track for new near-by radio-sources).
\end{enumerate}

\noindent The complete search is outlined in Algorithm~\ref{alg:ltt-ts}. It starts when the last 
spacecraft acquires its final orbit at the end of \emph{Phase 3a}, indicated by $\Delta t_0$. 

Lines 2--6 initialize a queue $Q$ with starting nodes for our search as described below. Firstly, the procedure \textsc{PropagateKeplerian} is used to compute the Keplerian celestial track $\mathcal P$ for an observation window of $\Delta t_w=30$ [days], starting from the time of the last observation shifted by the 15-day observation constraint (preventing any earlier observations). 
As there may be observations enforced in \emph{Phase 3a}, one has to ensure not to violate the 15-day constraint if the last observation of \emph{Phase 3a} was less than 15 days before $\Delta t_0$ (check omitted in the pseudocode). The procedure \textsc{FindPotentialTargets} finds all radio-sources along $\mathcal P$, which are within $\frac{1}{12}$ [rad] distance from the track. Note that observations with an estimated $J < 10^6$ [km] for $P=1$ and/or $\delta \ge 15.0$ are not considered, as we deem them to be suboptimal.
All other targets are stored in $Q$, which contains the spacecrafts orbits $\mathcal S$, the identifier of the next potential radio-source observation $n$, the approximate time of the observation $\Delta t_i^{-}$ and the sequence of previously observed radio-sources $V$. 

The information in $Q$ allows the search in lines 7--14 to update the trajectory by reevaluating potential new targets using the low-thrust targeting technique. Note that we do not consider observations that would extend the duration of \emph{Phase 3} by more than 15 moon periods, as we do need to provide enough time for \emph{Phase 2}
.
We use the Taylor integration (Line 10) to propagate the spacecraft states according to the vectors $u_1, u_2, u_3$ (shorthanded to just $u$ in the pseudocode) and $T^*$, as obtained from the distributed low-thrust targeting procedure. The next spacecraft states $S^\prime$, as well as the exact observation time $\Delta t_i^*$, are thus computed. The observation and thrust parameters are then concatenated with the current sequence of observations $V$ and stored for later retrieval and reproduction of the full trajectory.
At this point, we repeat the branching steps described before: a new Keplerian celestial track $\mathcal P$ is computed and, if the constraints allow it, new potential targets will be added to the queue $Q$. Once the designated 15 moon period 
is over or no potential targets are to be found
, $Q$ will eventually run out, and thus terminate the search.

\begin{figure}[ht]
\centering
\includegraphics[trim=100 100 40 100, width=0.32\columnwidth]{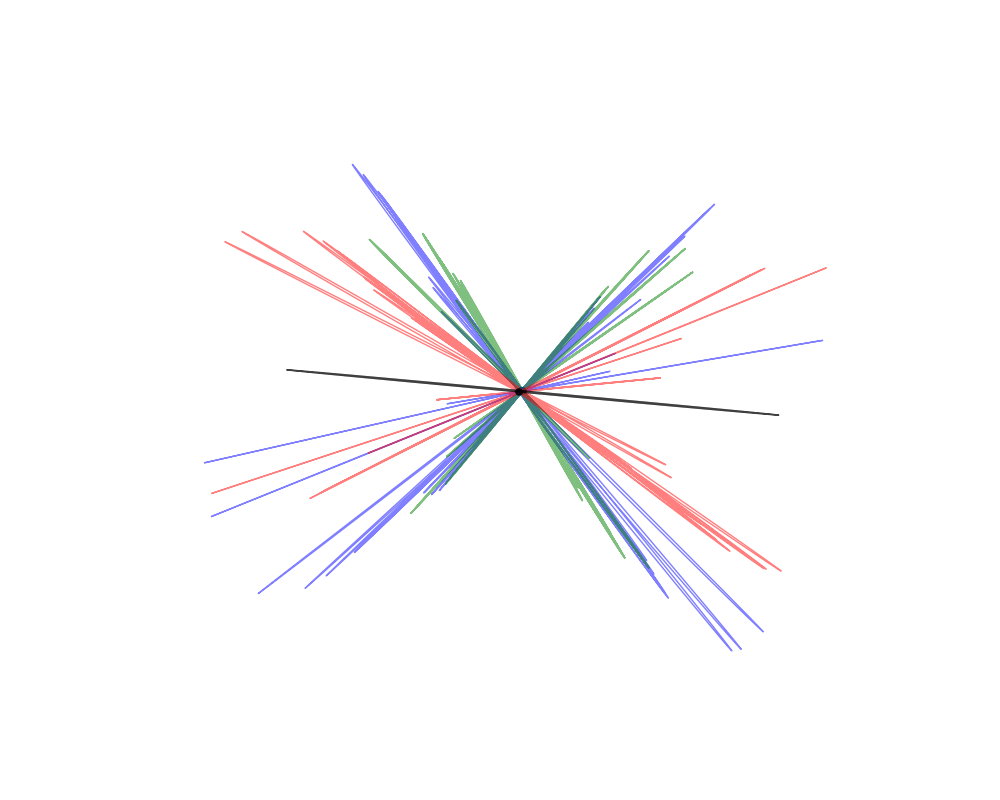}
\includegraphics[trim=120 100 60 100, width=0.32\columnwidth]{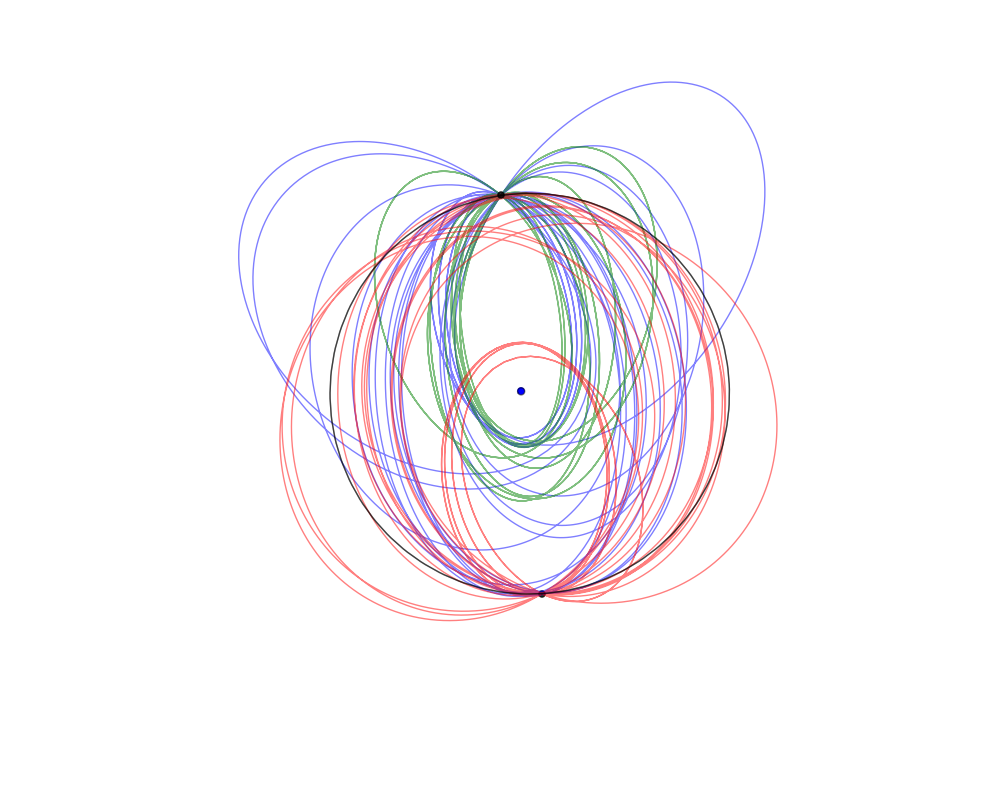}
\includegraphics[trim=100 100 100 100, width=0.32\columnwidth]{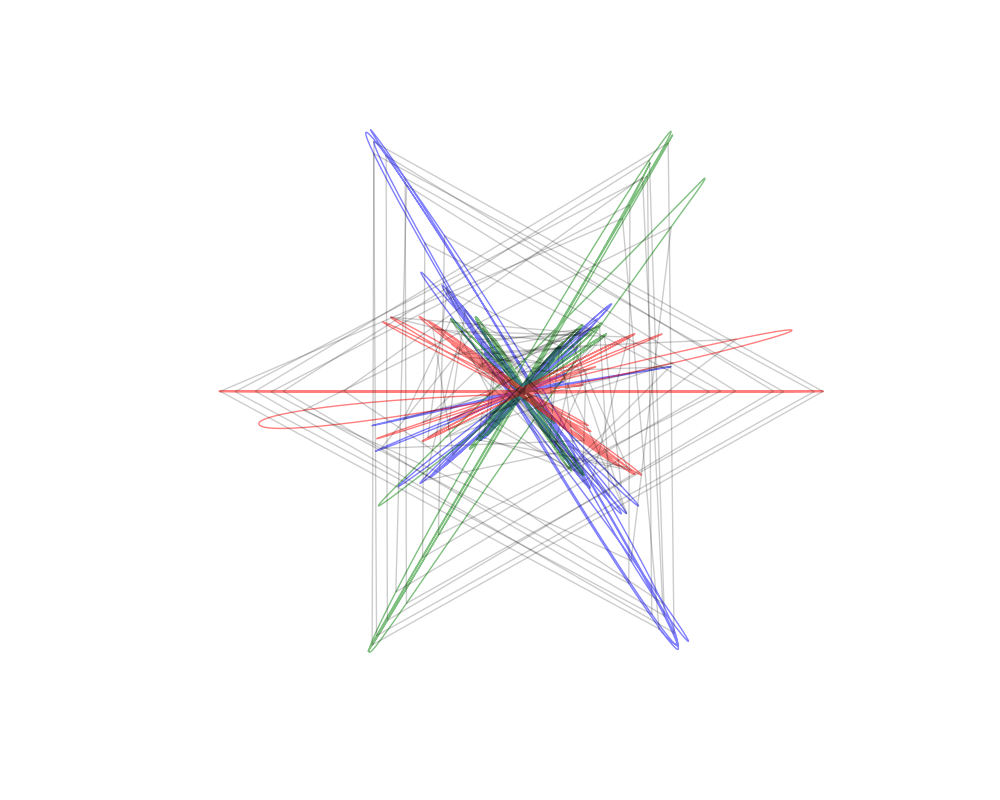}
\caption{Different views of \lq\lq The Nephalem\rq\rq. \emph{Phase 2} node view (left).  \emph{Phase 2} top view (center). Whole trajectory node view with observation triangles formed during the interferometric experiment (right). The three spacecarft trajectories are differentiated by color.\label{fig:nephalemview}}
\end{figure}

A key factor behind the success of this particular scheme is the idea that the initial ideal orbital configuration, as achieved by \emph{Phase 3a}, should be maintained as long as possible throughout the whole tree-search. Although the low-thrust targeting allows for the scoring of almost any target given enough time and thrust, a simple greedy approach would lead to a severe deviation from the ideal orbital configuration, resulting in significantly worse observation conditions for all future targets (e.g., by having lower altitudes). 
Keeping close to the ideal orbital geometry is achieved by a number of implementation details of the targetting procedure. Two key factors are the use of an objective function that minimizes cumulative thrust, and the small 
deviation constraint 
($\frac{1}{12}$ [rad] from the celestial track) 
for potential new observations. As visualized in Figure \ref{fig:nephalemview}, the orbits of the spacecraft 
remain very close to the ideal orbital geometry
throughout \emph{Phase 3}.

\begin{algorithm}[t!]
\caption{The low-thrust targeting tree-search.}
\begin{algorithmic}[1]
    \Function{LTT-tree-search}{$\mathcal S$, $\Delta t_0$, $\Delta t_w$}
    \State $Q = \emptyset$ \Comment{The first-in-first-out queue.}
    \State $V = \emptyset$ \Comment{Set of previously observed radio-sources. May also contain some elements from \emph{Phase 3a}}
    \State $\mathcal P = \;$\Call{PropagateKeplerian}{$\mathcal S$, $\Delta t_0 + 15$, $\Delta t_0 + \Delta t_w + 15$} \Comment{Computes Keplerian celestial-track.}
    \For {each $n$, $\Delta t_i^{-}$ in \Call{FindPotentialTargets}{$\mathcal P$}} \Comment{Loop over potential observations.}
        \State \Call{Q.Push}{$\mathcal S, n, \Delta t_i^{-}, V$} \Comment{Push the potential observation to queue.}
    \EndFor
    \While {$|Q| > 0$} \Comment{Continue as long as there are potential observations to process.}
    \State $\mathcal S, n, \Delta t_i^{-}, V = \;$\Call{Q.Pop()}{} \Comment{$n$ is a potential observation, $\Delta t^{-}_i$ an approx. observation time.}
    \State $u, T^* = \;$\Call{LT-Target}{$\mathcal S, n, \Delta t_i^{-}$} \Comment{Call the low-thrust targeting NLP solver.}
    \State $\mathcal S^\prime, \Delta t_i^{*} = \;$\Call{IntegrateTaylor}{$\mathcal S, u, T^*$} \Comment{New spacecrafts states $\mathcal S^\prime$, exact observation time $\Delta t_i^*$.}
    \State $V^+ = V \cup \{(n, \Delta t_i^{*}, u, T^*)\}$ \Comment{Extend the solution with a new observation.}
    \State $\mathcal P = \;$\Call{PropagateKeplerian}{$\mathcal S^\prime$, $\Delta t_i^{*} + 15$, $\Delta t_i^{*} + \Delta t_w + 15$} \Comment{Update the celestial-track.}
    \For {each $n+, \Delta t^{-}_i$ in \Call{FindPotentialTargets}{$\mathcal P$}} \Comment{Loop over potential observations.}
        \State \Call{Q.Push}{$\mathcal S^\prime, n^+, \Delta t_i^{-}, V^+$} \Comment{Add new potential observations for processing.}
    \EndFor
    \EndWhile
\State{\Return{$V^+$}} \Comment{$\mathcal O$ is assembled out of $V^+$.}
\EndFunction
\end{algorithmic}
\label{alg:ltt-ts}
\end{algorithm}

In Figure \ref{fig:lt_target}, several trajectories returned by the low-thrust targeting tree search are compared 
via three objectives:
mission value accumulated ($P=1$), number of observations made, and total \emph{Phase 3} duration.
Although the objective of having a high total $J$ during \emph{Phase 3} is clear, the remaining two objectives are also crucial, for they determine both the requirements and available time for \emph{Phase 2}.
Ideally we would like to have $P=6$ for all observations in \emph{Phase 3} but also to maximise the number of these observations.  However, if we do not restrict the number, there would simply be not enough time to score these observations with multipliers of $P=1$ and $P=3$ in \emph{Phase 2} (and \emph{Phase3a}).  We also must ensure that these targets are indeed observed in Phase 2 (some may not be feasible).  Therefore, we must trade-off a high $J$ against time and requirements for \emph{Phase 2} - requirements that are not yet determinable; hence, the single best trajectory of \emph{Phase 3} is undeterminable with certainty at this point.  For this reason several promising candidate trajectories are identified, by their potential scores,
after the termination of the tree search. 
The vectors of observations $\mathcal O$, observation times $\Delta t_i^*$ and baselines $h_\mathcal O$ are assembled out of all solutions stored in $V$. 
These candidates are then provided as part of the initial conditions for the design of the \emph{Phase 2} search.

%

\section{Phase 2 - Play the Moon}

The second phase of the trajectory is the last to be designed. From \emph{Phase 3}, a set of observed radio sources, $\mathcal O$, and associated baselines, $h_{\mathcal O}$, are known. 
\emph{Phase 1} provides the relative encounter velocities  $V_\infty^{\mathcal S_1}$, $V_\infty^{\mathcal S_2}$, $V_\infty^{\mathcal S_3}$ and the node to which they refer 
(e.g. [AN, AN, DN(+1)] indicates that $\mathcal S_1$ and $\mathcal S_2$ perform their first Moon fly-by at the ascending node while $\mathcal S_3$ at the following descending node). With this information available, at each successive Moon node passage, one or more fly-bys are designed whenever one or more spacecraft are present. In Figure \ref{fig:pullcrank}, we show a typical fly-by geometry; we also define the pull angle, $\alpha \in [0,\pi]$, and the crank angle, $k \in [0, 2\pi]$ \cite{Strange2007}. These angles allow one to compute the outgoing spacecraft velocity as:
$$
\mathbf v_{SC}^+ = \mathbf v_M + V_\infty \left( \sin\alpha\cos k \hat{\boldsymbol i}_n + \cos\alpha \hat{\boldsymbol i}_v  + \sin\alpha\sin k \hat{\boldsymbol i}_h \right),
$$
where, 
\begin{equation}
\hat{\boldsymbol i}_v = \frac{\mathbf v_{M}}{V_{M}}, \quad
\hat{\boldsymbol i}_h = \frac{\mathbf r_{M}}{r_{M}} \times \hat{\boldsymbol i}_v,  \quad
\hat{\boldsymbol i}_n = \hat{\boldsymbol i}_v \times \hat{\boldsymbol i}_h.
\end{equation}
At each fly-by, the incoming relative velocity $\mathbf v_\infty^-$ is assumed to be instantaneously rotated through an angle $\Delta$, hence, becoming $\mathbf v_\infty^+$, the outgoing relative velocity. Such a rotation angle cannot be larger than the value $\Delta_{max}$ computed as:
$$
\Delta_{max} = 2\arcsin\frac 1e, \qquad \qquad e^2 = \frac{r_m V_\infty^2 (r_m V_\infty^2 + 2\mu)}{\mu^2} + 1,
$$
so as not to exceed the required minimum Moon distance, $r_m$. In our case, such an angle ranges from 128\degg to 101\degg as $V_\infty$ ranges from $550$ to $900$ [m/s], indicating that most outgoing velocities are possible. In a first instance, the $\Delta_{max}$ constraint is thus neglected, and the pull and crank angles freely chosen. 

The pull angle, $\alpha$, alone determines the spacecraft velocity magnitude as $V_{SC}^{2} = V_\infty^{2} +V_M^2+2V_M V_\infty\cos\alpha$ and, therefore, also the orbit semi-major axis $a = \frac{r_M\mu}{2\mu-r_MV_{SC}^{2}}$ and period $\tau = 2\pi\sqrt{\frac{a^3}{\mu}}$. 
It is then chosen so to produce a resonant Moon orbit (thus ensuring one more fly-by will happen). The crank angle, $k$, 
controls the formation plane's orientation and so is selected based on a desired future observation; the technique we developed that allows such a choice is called $k$-targetting.
\input{pullcrank}
\subsection{$k$ - targeting}
To illustrate the basic idea behind the $k$-targeting, we discuss two separate cases. 
In both cases, given a list of target radio sources (e.g. $\mathcal O$), we find all 
crank angle values that will result in observing one of the radio-sources at 
$t\in[\underline t, \overline t]$ and compute the resulting baselines. 
\subsubsection{1-$k$ targeting }
\begin{figure}[t!]
\centering
\includegraphics[width=0.9\columnwidth]{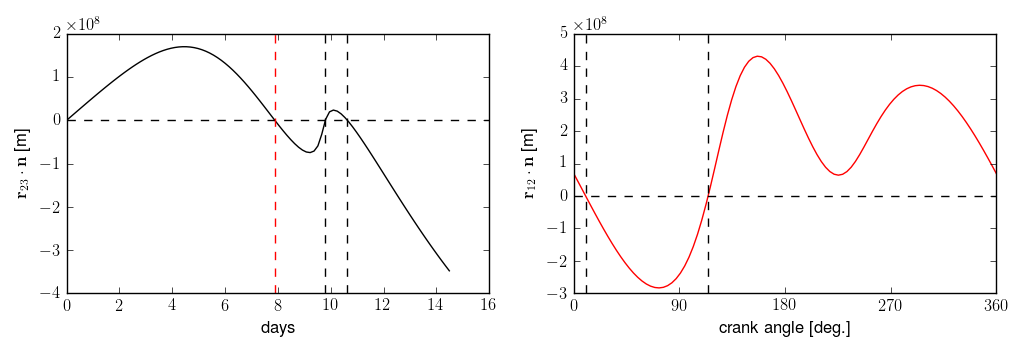}
\caption{1-$k$ targeting manoeuvre: multiple roots exist for both $\mathbf r_{23}(t) \cdot \mathbf n = 0$ and $\mathbf r_{12}(k) \cdot \mathbf n = 0$.}
\label{fig:1ktargeting}
\end{figure}

In a 1-$k$ targeting manoeuvre, two satellites $\mathcal S_2$ and $\mathcal S_3$ are on known orbits while the third satellite $\mathcal S_1$ performs a fly-by that will result in the observation of a radio-source $o\in \mathcal O$, with baseline $h$ at the epoch $t$. 
We consider a set $\mathcal T$ of possible periods containing at least $\tau_{1:1}, \tau_{1:2},\tau_{2:1},\tau_{3:2}$, where $\tau_{n:m} = \frac{n}{m}\tau_M$. 
We denote with $\mathbf n$ the direction of the radio-source $o$ and consider the equation:
\begin{equation}
\mathbf r_{23}(t) \cdot \mathbf n = 0,
\label{eq:r23}
\end{equation}
in the interval $[\underline t, \overline t]$, where $\mathbf r_{23} = \mathbf r_3 - \mathbf r_2$ is the relative position vector between $\mathcal S_3$ and $\mathcal S_2$ at $t$. Its solutions are indicated with $t_1, t_2, .. t_N$ and represent all the epochs in the chosen interval at which the radio-source $o$ can be observed, 
i.e., when $\mathcal S_2$ and $\mathcal S_3$ will have the correct alignment. To ensure that, for a fixed $t_i$, the position of $\mathcal S_1$ also allows for the 
observation, the crank angle $k$ must 
satisfy,
\begin{equation}
\mathbf r_{13}(k) \cdot \mathbf n = 0,
\label{eq:r13}
\end{equation}
where $\mathbf r_{13}(k) = \mathbf r_3-\mathbf r_1(k)$ is the relative position vector between $\mathcal S_3$ and $\mathcal S_1$  computed at $t_i$.  This clearly depends on the crank angle chosen to perform the $\mathcal S_1$ fly-by. As shown in Figure \ref{fig:1ktargeting},
both equations above are well behaved 
and easily solved with a zero finding routine.  Algorithm \ref{alg:1k} was used to compute all possible 1-$k$ targeting manoeuvres.

\begin{algorithm}[t!]
\caption{The 1-$k$ targeting algorithm}
\begin{algorithmic}[1]
\Function{1k-Targeting}{$\mathcal O$, $\mathcal T$, $\underline t$, $\overline t$}
\State{$\mathcal{M} \gets \emptyset$}
\ForAll{$o \in \mathcal O$} \Comment{consider all target radio-source}
\ForAll{$\tau \in \mathcal T$} \Comment{consider all possible resonance}
\State find $\mathbf t = [t_1, .. t_N]$ , i.e. all roots in $[\underline t, \overline t]$ of $\mathbf r_{23}(t) \cdot \mathbf n = 0$ 
\ForAll{$t \in \mathbf t$}
\State find $\mathbf k = [k_1, .. k_M]$ , i.e. all roots of $\mathbf r_{13}(k) \cdot \mathbf n = 0$
\ForAll{$k \in \mathbf k$}
\State $\mathcal{M} \gets \mathcal{M} \cup \{((\tau, k), t, h, o)\}$
\EndFor
\EndFor
\EndFor
\EndFor
\State{\Return{$\mathcal{M}$}}
\EndFunction
\end{algorithmic}
\label{alg:1k}
\end{algorithm}

\begin{algorithm}[t!]
\caption{The 2-$k$ targeting algorithm}
\begin{algorithmic}[1]
\Function{2k-Targeting}{$\mathcal O$, $\mathcal T$, $\mathbf t$}
\State{$\mathcal{M} \gets \emptyset$}
\ForAll{$o \in \mathcal O$} \Comment{consider all target radio-source}
\ForAll{$\tau \in \mathcal T$} \Comment{consider all possible resonance}
\ForAll{$t \in \mathbf t$}
\State find $\mathbf k = [k_1, .. k_M]$, i.e. all roots of $\mathbf r_{13}(k) \cdot \mathbf n = 0$
\State find $\tilde{\mathbf k} = [\tilde k_1, .. \tilde k_M]$, i.e. all roots of $\mathbf r_{23}(k) \cdot \mathbf n = 0$
\ForAll{$k \in \mathbf k$}
\ForAll{$\tilde k \in \tilde{\mathbf k}$}
\State{$\mathcal{M} \gets \mathcal{M} \cup \{((\tau, k), (\tau, \tilde k), t, h, o)\}$}
\EndFor
\EndFor
\EndFor
\EndFor
\EndFor
\EndFunction
\State{\Return{$\mathcal{M}$}}
\end{algorithmic}
\label{alg:2k}
\end{algorithm}

\subsubsection{2-$k$ targeting }
In a 2-$k$ targeting manoeuvre, one satellite (e.g. $\mathcal S_3$) is on a known orbit while the other satellites (e.g. $\mathcal S_1$ and $\mathcal S_2$) perform fly-bys that will result in the observation of a radio-source $o\in \mathcal O$, with baseline $h$ at the epoch $t$. As opposed to the 1-$k$ case, 
$t$ 
cannot be restricted by the dynamics of the uncontrolled satellites, 
for in this case, there is only one, 
$\mathcal S_3$. We may then consider 
$t$ as a free parameter necessary to compute the position $\mathbf r_3(t)$ of $\mathcal S_3$ and find the crank angles of the two fly-by manoeuvres by solving 
$\mathbf r_{13}(k) \cdot \mathbf n = 0$ and $\mathbf r_{23}(k) \cdot \mathbf n = 0$ for $k$. The corresponding algorithm is reported as Algorithm \ref{alg:2k} and assumes that a list of epochs $\mathbf t$ is given.

\subsection{The inner tour tree search}
\emph{Phase~2} is designed last; thus, a number of (patching-)constraints need to be fulfilled during its design to obtain an overall valid mission.
The node arrival conditions from \emph{Phase~1} result in temporal and spatial constrains as well as 
the fly-by patching constraint on $\Delta_{max}$.
The \emph{Phase~2} trajectory for a particular spacecraft can not start before the node arrival.  It has to start at the corresponding AN or DN node, and the fly-by patching conditions have to be met. 
\emph{Phase~3} also sets similar temporal, spatial and fly-by patching conditions that must be satisfied. 

In addition to the above constraints, the set of observed radio sources $\mathcal O$ and their baselines $h_{\mathcal O}$ are known from the design of \emph{Phase~3}.
A straightforward strategy, therefore, is to plan \emph{Phase~2} so that all of these radio sources are observed twice, 
with a suitably increased baseline $h$ on each second observation, ensuring a $P=6$ observation in \emph{Phase~3}.
We refer to the baselines of three successive observations of the same radio source as $h_1, h_2$ and $h_3$.  Assuming an increasing baseline, to successfully score $P=3$ and $P=6$ multipliers on the second and third observations respectively, the problem description\cite{gtoc8description} demands $9 h_1 \leq 3 h_2 \leq h_3$. 
Hitting these bounds exactly is infeasible.


Initial results showed that the solution to this constrained optimization problem is 
more complex then simply targeting all sources sequentially: at a low altitude with multiplier $P=1$ 
and subsequently at a higher altitude to reach a multiplier $P=3$. 
We can employ a greedy selection that considers all possible k-maneuvers at the current state and evaluates 
the resulting list of possible observations, incrementally picking the best maneuver with respect 
to the final objective $J$ (including changes in multipliers).  This, however, results in a suboptimal design that 
fails to increase the multipliers of nearly all 
observations to $P=6$ in \emph{Phase~3}.


An alternative is to build a solution in reverse, starting at the initial conditions for \emph{Phase~3} and working backwards
to meet the final conditions of \emph{Phase~1}. In this way, the first maneuver to be picked determines the last observation to be made before
the spacecraft embark on the leg to the outer orbits (\emph{Phase~3a}).


To address this complex sequential decision making problem, we use a heuristic search method called \emph{beam search}.
Beam search works like breadth-first search in that it explores the tree in layers of increasing depth; however, only 
the highest-ranked nodes of each layer are branched. A pseudo code version 
is listed in Algorithm~\ref{alg:search-daniel}.
Starting from an initial state $\mathbf{s}$ that describes the patching conditions to the start of \emph{Phase~3}, a search frontier
$\mathcal{F}$ is built by computing all possible successors of $s$ (see {\sc Branch}). This function recognises whether there are one or two satellites at the considered moon node. The applicable temporal constraints are evaluated, and all corresponding 
1-$k$ or 2-$k$ targeting maneuvers are considered. 
When a 
radio source is targeted for the first time in \emph{Phase~2}, an 
altitude above a lower bound is enforced:
$h_2 > (\frac{1-\epsilon}{3})h_3$ where $\epsilon$ is a tolerance set as one of the search parameters.
As the goal is to target all sources twice during \emph{Phase~2} (and we search backwards), this 
observation
will reach a $P = 3$ multiplier and, thus, contribute significantly to the final score. An additional benefit is that this allows for a looser
upper bound on the altitude of the $P=1$ observation of the same source. All necessary patching conditions, either across phases or
between moon fly-bys, are evaluated. 
If successful, the maneuver is concatenated with the current state, and the result is added to 
the list of successors.

\begin{algorithm}[t!]
\caption{Beam search}
\begin{algorithmic}[1]
\Function{Branch}{$\mathbf s$, $\mathcal{O}$}
\State{$\mathcal{S} \gets \emptyset$}
\Comment{list of successor states}
\State{$\mathcal{O}' \gets \mathcal{O} \setminus \{o \in O\ |\ \mbox{$o$ observed twice in $\mathbf s$}\}$}
\State{$\mathcal{T} = \{\frac{1}{2}, 1, \ldots\}$}
\If{\Call{AtNode}{$\mathbf s$} = 1}
\Comment{only one satellite is at the node}
\State{${\underline t}, {\overline t} \gets $ \Call{1k-Bounds}{$\mathbf s$}}
\ForAll{$((\tau, k), t, h, o) \in $ \Call{1k-Targeting}{$\mathcal{O}'$, $\mathcal T$, $\underline t$, $\overline t$}}
\If{$o$ not observed in $\mathbf s$ \textbf{and} $h > (\frac{1-\epsilon}{3})h'$}
\Comment{where $h'$ is the altitude of $o$ in \emph{Phase~3}}
\State{\textbf{continue}}
\EndIf
\If{\Call{Patch}{($\tau$, $k$), $\mathbf s$}}
\State{$\mathcal{S} \gets \mathcal{S} \cup \{((\tau, k), t, h, o) \oplus \mathbf{s}\}$}
\EndIf
\EndFor
\Else
\Comment{two satellites are at the node}
\State{${\mathbf t} \gets $ \Call{2k-Bounds}{$\mathbf s$}}
\ForAll{$((\tau, k), (\tau, \tilde{k}), t, h, o) \in $ \Call{2k-Targeting}{$\mathcal{O}'$, $\mathcal T$, $\mathbf t$}}
\If{$o$ not observed in $\mathbf s$ \textbf{and} $h > (\frac{1-\epsilon}{3})h'$}
\State{\textbf{continue}}
\EndIf
\If{\Call{Patch}{($\tau$, $k$), ($\tau$, $\tilde k$), $\mathbf s$}}
\State{$\mathcal{S} \gets \mathcal{S} \cup \{((\tau, k), (\tau, \tilde{k}), t, h, o) \oplus \mathbf s\}$}
\EndIf
\EndFor
\EndIf
\State{\Return{$\mathcal{S}$}}
\EndFunction
\Statex{}
\Function{Search}{$\mathbf s$, $bs$, $\mathcal O$}
\State{best $\gets$ NULL}
\State{$\mathcal{F}\gets \{\mathbf{s}\}$}
\Comment{initialize search frontier with starting state $\mathbf s$}
\While{$|\mathcal{F}| > 0$}
\State{$\mathcal{F'}\gets \emptyset$}
\ForAll{$\mathbf s \in \mathcal{F}$}
\State{$\mathcal{F}'\gets \mathcal{F}'\ \cup $ \Call{Branch}{$\mathbf s$, $\mathcal{O}$}}
\EndFor
\State{$\mathcal{F} \gets \emptyset$}
\State{best $\gets \argmax_{\mathbf{s}' \in \mathcal{F'}} $ \Call{Score}{$\mathbf{s}'$, $\mathcal{O}$}}
\While{$\mathcal{F'} \neq \emptyset$ \textbf{and} $|\mathcal{F}| < bs$}
\Comment{reduce to beam size}
\State{$\mathbf{s}\gets \argmax_{\mathbf{s}' \in \mathcal{F'}} $ \Call{Rank}{$\mathbf{s}'$, $\mathcal{O}$}}
\State{$\mathcal{F}' \gets \mathcal{F}' \setminus \{\mathbf{s}\}$}
\State{$\mathcal{F} \gets \mathcal{F} \cup \{\mathbf{s}\}$}
\EndWhile
\EndWhile
\State{\Return{best}}
\EndFunction
\end{algorithmic}
\label{alg:search-daniel}
\end{algorithm}

For each state in the frontier, all successors are added to a temporary new frontier $F'$.  The highest scoring state is evaluated by 
calculating $J$ using the list of observations in $\mathbf{s}$ and in $\mathcal O$.
Only the highest-ranked nodes of the temporary frontier $F'$ are considered in the next iteration, thus reducing the beam size to $bs$.  
We experimented with different heuristics for the {\sc Rank} function:
\begin{itemize}
    \item \textbf{$J$ optimal.} States are ranked by score $J$ in decreasing order. 
    \item \textbf{Time average.} States are ranked by the ratio of time consumed for observations to total time in \emph{Phase~2}: $\frac{15 [days] * n}{t}$.
    \item \textbf{Time optimal.} States are ranked by time consumed in \emph{Phase~2} in increasing order.   
\end{itemize}
The search resulting in the winning trajectory used a rather `aggressive' score pruning with $\epsilon = 0.25$ combined with average time ranking. 


\section{Results}

In order to identify the different trajectories and ideas studied during the competition timeframe, we used the naming theme of video-games. The best trajectory found, named \lq\lq The Nephalem\rq\rq\ after the main hero of the Diablo video-games, has a primary objective $J=146.33$ [Mkm] and makes $n=45$ observations of $m=17$ distinct radio sources. Several alternative trajectory options were found making use of different spirals and tree searches. 
Their scores ranged from $125.3-145.26$ [Mkm], indicating that the solution space is rather dense 
and, hence, \lq\lq The Nephalem\rq\rq\ is not an isolated opportunity.

\subsection{The Nephalem}
\begin{figure}[t]
\centering
\includegraphics[width=0.9\columnwidth]{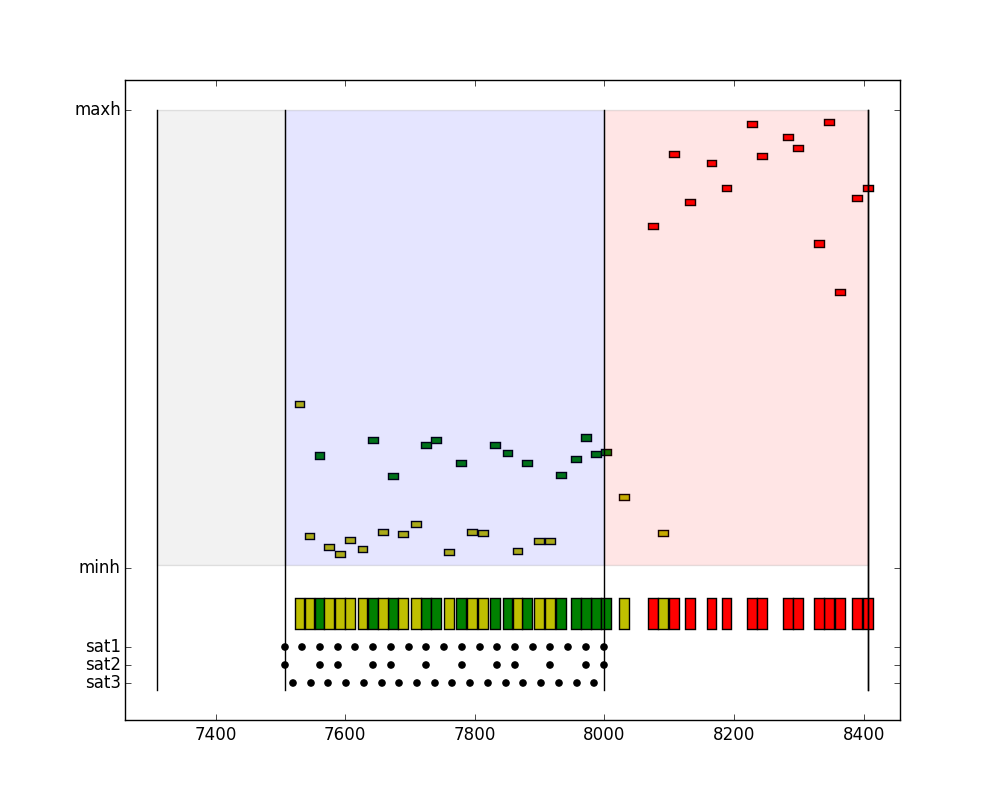}
\caption{Observation schedule of \lq\lq The Nephalem\rq\rq. Phase 1, Phase 2 and Phase 3 are clearly indicated as gray, blue and red shaded rectangles respectively. The height of each observation is reported in the y-axis as well as the P value (yellow, green and red color for 1, 3 and 6 respectively). The width of the bars correspond to 15 days (no overlaps are thus possible). The nodes at which a fly-by targeting manoeuvre is performed are also visualized separately for the three satellites.}
\label{fig:schedule}
\end{figure}

We send two spacecraft, say $\mathcal S_1$ and $\mathcal S_2$,  to the moon ascending node with a final $v_{\infty} = 866$ [m/s] and one, say $\mathcal S_3$, to the Moon's descending node with a final $v_{\infty} = 555$ [m/s]. The descending node spiral was generated using the tangential thrust control law.  The other spiral, assumed to be the same for both $\mathcal S_1$ and $\mathcal S_2$, was generated using the inversion control law with $\nu = 3.75$ \degg. The higher $v_\infty$ were selected 
to allow for the acquisition of the high inclination orbits of the ``ideal'' geometry in a short time frame during \emph{Phase 3a}. The third spacecraft, $\mathcal S_1$ and $\mathcal S_2$, will only need to acquire a zero inclination orbit. 
It can, therefore, benefit from a greater flexibility when targeting observations during \emph{Phase 2} via smaller values of $v_\infty$. Starting from the designed spirals, we construct \emph{Phase 3a} so that the three spacecraft are injected into circular orbits having $a = 999000$ [km], $e=0$ and $i=0, 60, 60$ \degg (the RAAN of the 60 \degg orbits being 180 \degg apart). We also target a few interferometric observations during the transfer. We then start \emph{Phase 3b}: a sequence of low-thrust targeting manoeuvres that end with the last observation being taken after $tof = 2.98$ [years] and a list $\mathcal O$, containing 16 observed radio-sources. 
Fourteen of these have a high baseline, $h$, while two, taken immediately after the last Moon encounter have a low $h$ and, thus, will not be high-priority during our \emph{Phase 2} design. 

\begin{figure}[t]
\centering
\includegraphics[width=0.9\columnwidth]{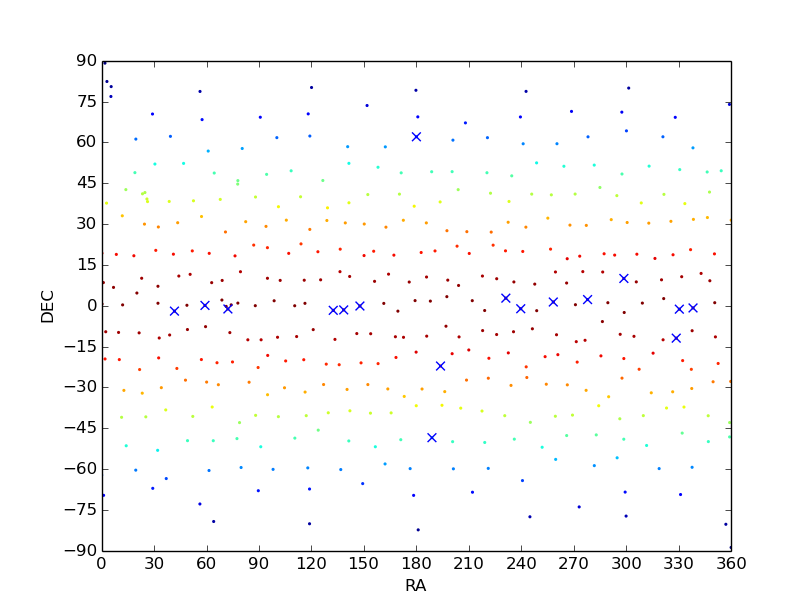}
\caption{Map of the radiosources observed by the \lq\lq The Nephalem\rq\rq. The high 
declination observations correspond to those inserted at the interface between phases and
only observed once ($P=1$).\label{fig:observed}}
\end{figure}

After the design of \emph{Phase 2}, we reach a trajectory value of $J=146.33$ [Mkm]. The overall trajectory is summarized in the plot reported in Figure \ref{fig:schedule} where the observation schedule of \lq\lq The Nephalem\rq\rq\ is depicted. 
The nodes at which a fly-by targeting manoeuvre is performed during \emph{Phase 2} are also visualized separately for the three satellites at the bottom of the figure. 
$\mathcal S_3$ performs a 1-$k$ targeting manoeuvre at each node, while  $\mathcal S_1$ performs a 1-$k$ or a 2-$k$ targeting manoeuvre according to the presence of $\mathcal S_2$. 
All the high $h$ radio-source observations in $\mathcal O$ are observed with a $P=6$ multiplier, and the average number of days between observations in \emph{Phase 2} is 17.07 (thus quite close to the 15 [days] prescribed), indicating a good performance of the tree-search in 
\emph{Phase 2}. The average number of days between observations in \emph{Phase 3} is 24.37. In the orbital geometry targeted, the ideal satellite positioning only happens roughly every 55 [days]; hence, 24.37 [days] between good observations is deemed as a good frequency. 
As shown in Figure \ref{fig:schedule}, three observations are made only once and, thus, do not contribute much to the overall score. These observations were inserted opportunistically and not as a result of the tree searches. The first one is at the interface of \emph{Phase 1} and \emph{Phase 2}, when $\mathcal S_3$ is still spiralling out to the Moon but $\mathcal S_2$ and $\mathcal S_3$ have already made their first Moon fly-by. They can, therefore, perform a 2-$k$ targeting manoeuvre. The other two are the result of enforcing path constraints during \emph{Phase 3a} and have a rather high declination, as shown in Figure \ref{fig:observed}, which prevents them from being selected as good candidates for a repeated observation in \emph{Phase 2}.

\section{Alternative design strategies}
While the main design strategy described revealed to be quite successful, other mission options were also considered to be competitive. In particular, one can allow the Moon to be encountered at three distinct positions along its orbit allowing the three spacecraft 
to perform a 1-$k$ targeting manoeuvre with a higher frequency. The potential advantage would be to reduce the average number of days between observations in \emph{Phase 2}, at the cost of lengthening the spiralling out phase.
A second option would be to design \emph{Phase 2} without making systematic use of Moon fly-bys and using only thrust to target radio-sources. We refer to this option as low-thrust formation flying. A third option is to not reach the 1,000,000 [km] ideal orbital configuration with all three spacecraft but to continue scoring observations using the $k$-targeting manoeuvres until the end. This could potentially result in more $P=6$ observations taken at the cost of a smaller $h$. All the above strategies were investigated briefly and discarded as soon as a first indication on the potential final score for the chosen baseline (i.e. sending all the three spacecraft to reach the 1,000,000 [km] after multiple Moon fly-bys) was grasped.

\subsection{Phase 2-3 alternative strategy: Pure Moon resonance}

In a Moon resonance strategy, all observations ($P=1,3,6$) are targeted using lunar fly-bys and resonant orbits. This strategy was preliminary assessed by considering some orbital configuration and computing its potential single observation score (the maximum single-observation score among all possible celestial tracks), the average frequency of observations,
and the global coverage. Table \ref{tab:MoonResonance} lists some examples, with a short description and the
maximum score for a single observation $J_{i}$. As it seemed possible to reach decent single observation $J_i$ scores with this alternative set-up, the tree search used during \emph{Phase 2} was run and developed also towards designing the full trajectory without a \emph{Phase 3}. Pursuing this concept, trajectories having $J=90$ [Mkm] were found and a crude estimate of a potential maximum of $J=120$ [Mkm] was possible, which advocated 
the addition of \emph{Phase 3}.

\begin{table}
\caption{Moon resonance strategies.\label{tab:MoonResonance}}
\noindent \centering{}%
\begin{tabular}{l|l|l}
\hline 
Name & Description & $J_{i}$\\
\hline 
\multirow{2}{*}{Nibbler} & $\mathcal{S}_{1}$, $\mathcal{S}_{2}$ on inclined, 1e6km-radius  & \multirow{2}{*}{1.1e6 km}\\
 & circular orbit; $\mathcal{S}_{3}$ in 1:1 resonance. & \\
\multirow{2}{*}{Tetris} & $\mathcal{S}_{1}$ on 32 deg , 1e6km-radius circular orbit;  & \multirow{2}{*}{0.9e6 km}\\
 & $\mathcal{S}_{2}$, $\mathcal{S}_{3}$ in 1:1 resonance at opposite
nodes. & \\
\multirow{2}{*}{Pong} & $\mathcal{S}_{1}$ remain on low-earth orbit;  & \multirow{2}{*}{1.0e6 km}\\
 & $\mathcal{S}_{2}$, $\mathcal{S}_{3}$ in 2:1 resonance. & \\
Double Dragon & $\mathcal{S}_{1}$ in 1:1 resonance; $\mathcal{S}_{2}$, $\mathcal{S}_{3}$
in 2:1 resonance. & 1.2e6 km\\
\hline 
\end{tabular}
\end{table}

\begin{figure}[ht!]
\noindent \begin{centering}
\includegraphics[width=10cm]{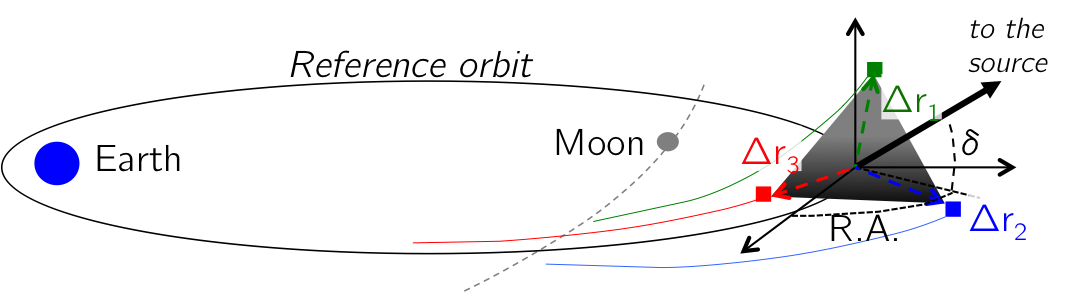}
\par\end{centering}
\caption{Schematic of the \emph{low-thrust formation flying} strategy (phase
2) during one observation. The three spacecraft orbits are visualized in green, red and blue.\label{fig:LTFF scheme}}
\end{figure}

\begin{figure}[t!]
\noindent \begin{centering}
\includegraphics[width=10cm]{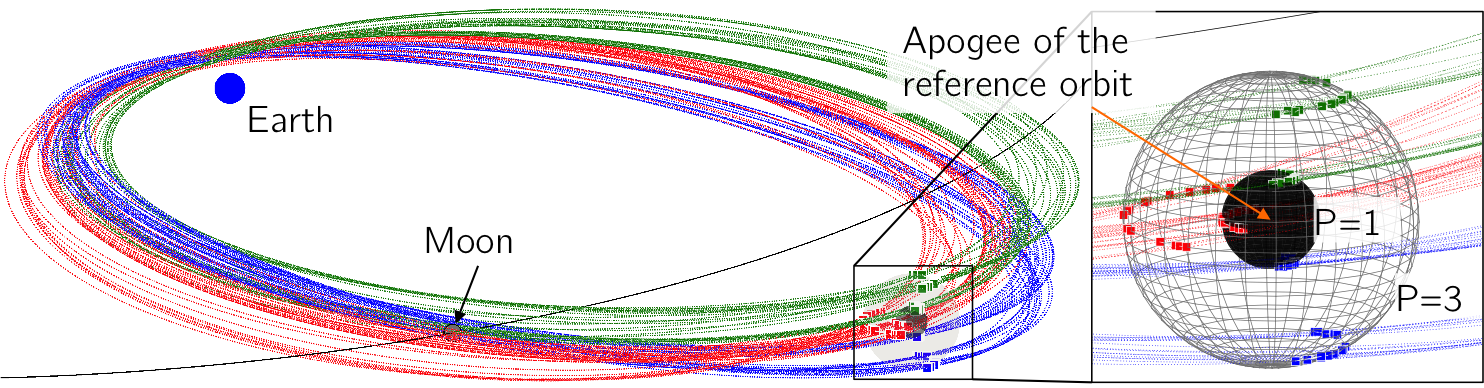}
\par\end{centering}
\caption{Example solution of the \emph{low-thrust formation flying} strategy
with 14 sources, observed twice.  Trajectories of the three spacecraft are illustrated in blue, green and red; while, their locations at the times of the observations are marked with squares. The spheres of radius $2h/3$ with $h=(10,000\;30,000)$ are also shown \label{fig:LTFF_sol}.}
\end{figure}

\subsection{Phase 2 alternative strategy: low-thrust formation flying }

Using low-thrust reachable set theory \cite{Campagnola2015}, one can estimate that the orbit control capabilities of $\mathcal S_{i}$ on a highly-eccentric 15-day orbit are sufficiently large to displace the spacecraft to any location on a 20,000 [km] sphere around its apogee. The low-thrust formation flying strategy was thus developed; here, $\mathcal{S}_{1}$, $\mathcal{S}_{2}$ and $\mathcal{S}_{3}$ fly in close formation and perform one observation close to the apogee of the reference orbit. Given the list $\mathcal O$ of radio sources from \emph{Phase 3}, we can try to observe them all exactly twice by targeting 
a triangular formation at each orbit apogee, initially with $h=10,000$ [km] and later with $h=30,000$ [km].

For all spacecraft, such a phase would start and end with a lunar fly-by; all fly-bys occur at the same lunar node with the same $V_{\infty}$. In between, each spacecraft perform $2|\mathcal{O}|+1$
revolutions: one revolution for each $h=10,000$ [km] observation,
another for each $h=30,000$ [km] observation, and one
final revolution to return to the Moon. Figure \ref{fig:LTFF scheme}
shows a schematic of such a strategy, visualizing the three spacecraft on an equilateral
triangle formation at the time of the observation, and the reference
orbit passing through the center of the triangle at apogee. Given
a desired radio-source, $o$, with declination $\delta$ and right ascension $RA$, low-thrust arcs are used to place the three spacecraft $\mathcal{S}_{i}$ at the relative positions:
\[
\Delta r_{i}=\frac{2}{3}h\left(\begin{array}{c}
-\sin RA\cos f_{i}-\cos RA\sin\delta\sin f_{i}\\
\cos RA\cos f_{i}-\sin\delta\sin RA\sin f_{i}\\
\cos\delta\sin f_{i}
\end{array}\right),\quad f_{i}=\left(\frac{\pi}{3},-\frac{\pi}{3},\pi\right).
\]
Figure \ref{fig:LTFF_sol} shows one of the solutions obtained with $|\mathcal{O}|=14$. 
For each spacecraft, an initial guess trajectory is assembled, revolution by revolution, from ballistic arcs centered at the observation locations. The entire trajectory is then optimized with a direct method\cite{Campagnola2015}, using low-thrust arcs and the initial Moon fly-by to restore continuity between consecutive arcs.

Overall, using this strategy is always possible to observe all the sources in $\mathcal{O}$ twice (with $P=1$ and $P=3$), using $n=\mbox{ceil}\left(15\frac{2|\mathcal{O}|+1}{27.3}\right)$ Moon revolutions. In the example solution of Figure \ref{fig:LTFF_sol}, the time of flight is $16$ [lunar months], and the average frequency of observation is 15.6 [days], just above the 15 [days] constraint. Despite the high frequency of observations, this strategy was abandoned for several reasons. Firstly, 
while contributing to only a fraction of the final score (maximum 20\%),  the $J$ score of the $P=1$ and $P=3$ observations is better optimized using the \emph{Play the Moon} approach. Secondly, the most promising \emph{Phase 3a} trajectories that we found started with the spacecraft at different nodes, and with different $V_{\infty}$. In order to reach these conditions at the end of a low-thrust formation flying phase, at least 1.5 [lunar months] should be added to the duration. In fact, one spacecraft has to be at the opposite node with the correct $V_{\infty}$, thus decreasing the average frequency of observations. One last issue with this strategy, though deemed as minor, is that it makes use of propellant mass, which is a secondary objective in this mission.

\section{Conclusions}
The design of an interplanetary trajectory in response to the problem formulated in the 8th edition of the Global Trajectory Optimization Competition involved the development of a number of new techniques that are of general interest to mission design. The Moon targeting technique used to design \emph{Phase 1} maps the problem of designing a two-dimensional spiral to that of a one-dimensional zero-finding problem while allowing for flexibility to design the arrival $V_\infty$. The 1-$k$ and 2-$k$ targeting manoeuvres used for the design of \emph{Phase 2} allow one to achieve exact orientations of the satellite formation plane, searching efficiently in the crank angle space. The distributed low-thrust targeting technique used for the design of \emph{Phase 3b} achieves the same goal, making optimal use of a distributed propulsion system. 
Evidently, the mission design for the \lq\lq high-resolution mapping of radio sources in the universe using space-based Very-Long-Baseline Interferometry (VLBI)\rq\rq, resulted in a rather complex trajectory where advanced algorithmic solutions were developed that are still subject of our active research.

\bibliographystyle{AAS_publication}   
\bibliography{ref}

\end{document}

%% file: idealgeometry.tex
\begin{figure}[t!]
\begin{center}
\begin{tikzpicture}[thick]
    \pgfmathsetmacro{\R}{3}
    
    \path[draw, thin, dashed] ({\R*cos(180)},{\R*sin(180)}) coordinate [label= left:$\color{blue} \mathcal S_1$] (A)
            -- ({\R*cos(60)},{\R*sin(60)}) coordinate [label=above:$\color{green} \mathcal S_2$] (C)
            -- ({\R*cos(-60)},{\R*sin(-60)}) coordinate [label=below:$\color{red} \mathcal S_3$] (B)
            -- cycle;
            
    \fill [blue] (A) circle (2pt);
    \fill [green] (C) circle (2pt);
    \fill [red] (B) circle (2pt);
    \draw [color=gray] circle(\R);
    
    \node (A1) at ({-\R*cos(180)},{-\R*sin(180)}) {};
    \node (C1) at ({-\R*cos(60)},{-\R*sin(60)}) {};
    \node (B1) at ({-\R*cos(-60)},{-\R*sin(-60)}) {};
    \node (A2) at ({\R*cos(180)*0.7},{\R*sin(180)*0.7}) {};
    \node (C2) at ({\R*cos(60)*0.7},{\R*sin(60)*0.7}) {};
    \node (B2) at ({\R*cos(-60)*0.7},{\R*sin(-60)*0.7}) {};
    \path[draw, thin, blue] [] (A) -- (A1);
    \path[->,draw, , blue] [] (A) -- (A2);
    \path[draw, thin, red] [] (B) -- (B1);
    \path[->,draw, , red] [] (B) -- (B2);
    \path[draw, thin, green] [] (C) -- (C1);
    \path[->,draw, , green] [] (C) -- (C2);

    \draw[thin, gray] plot[domain=-\R*0.3:\R*0.3] (\x, -0.1*\x);
    
    \draw[thin, <->] (-5,0) -- (-5, \R) node[text depth=5ex,rotate=90] {$1,000,000$ Km};
\end{tikzpicture}
\begin{tikzpicture}[thick]
    \pgfmathsetmacro{\R}{3}
    
    \path[draw, thin, dashed] ({\R*cos(180)},{\R*sin(180)}) coordinate [label= left:$\color{blue} \mathcal S_1$] (A)
            -- ({\R*cos(60)},{\R*sin(60)}) coordinate [label=above:$\color{green} \mathcal S_2$] (C)
            -- ({\R*cos(-60)},{\R*sin(-60)}) coordinate [label=below:$\color{red} \mathcal S_3$] (B)
            -- cycle;
            
    \fill [blue] (A) circle (2pt);
    \fill [green] (C) circle (2pt);
    \fill [red] (B) circle (2pt);
    \draw [color=gray] circle(\R);
    
    \draw [->, blue] (A) arc (180:190:4) ;
    \draw [->, red] (B) arc (-60:-50:4) ;
    \draw [->, green] (C) arc (60:70:4) ;

    \draw[thin, gray] plot[domain=-\R*0.3:\R*0.3] (\x, -0.1*\x);
    
\end{tikzpicture}
\end{center}
\caption{Ideal orbital geometries observed from an ideally placed
radio source on the Moon line of nodes. The 0$^o$, 60$^o$, 60$^o$ (left) and the 90$^o$, 90$^o$, 90$^o$ (right) are visualized.\label{fig:ideal}}
\end{figure}

%% file: spirals.tex
\begin{figure}[t!]
\begin{center}
\begin{tikzpicture}[scale=0.5, dot/.style={circle,inner sep=1pt,fill,label={#1},name=#1},
  extended line/.style={shorten >=-#1},
  extended line/.default=1cm]
    \pgfmathsetmacro{\ra}{5}
    \pgfmathsetmacro{\rp}{1}

    \draw[samples=500, dashed, red] plot[domain=0:1250] ({-(2*(\ra*(1+\x/360))*\rp) / ((\ra*(1+\x/360))+\rp) / (1 + ((\ra*(1+\x/360)) - \rp) / ((\ra*(1+\x/360)) + \rp)* cos(\x)) * cos(\x)}, {-(2*(\ra*(1+\x/360))*\rp) / ((\ra*(1+\x/360))+\rp) / (1 + ((\ra*(1+\x/360)) - \rp) / ((\ra*(1+\x/360)) + \rp) * cos(\x)) * sin(\x)});
    
    \draw[gray] (17.15,-6.5) arc (-40:40:10cm);
    \draw[fill=gray] (18.9,-3.39) circle (0.2cm) node[below] {$\mbox{target node}$};   
    
    \draw[gray] (0,0) circle (1cm);
    \draw[fill=gray] (0.34,0.939) circle (0.2cm);

    \draw[thin, extended line=1.5cm] (0,0) -- (0.34,0.939);   
    \draw[thin] (0,0) -- (18.9,-3.39) node[above, midway, color=black, rotate=-10] {line of nodes};
    \draw[thin, extended line=1.5cm] (0,0) -- (-1,0);
    
    \draw[<->] (0.68404028665,1.87938524157) arc (70:-10:2cm) node[midway, above] {$\theta_1$};   
    \draw[<->] (-2,0) arc (180:350:2cm) node[midway, below] {$\theta_2$};

\end{tikzpicture}
\end{center}
\caption{Spiral design is obtained by finding the thrust $T$ as a solution to the equation $\theta_1(T)+\theta_2(T) = 2\pi$. \label{fig:spiral}}
\end{figure}

%% file: pullcrank.tex
\begin{figure}[t!]
\begin{center}
\tikzstyle{marknode}=[circle, fill=black, inner sep=0pt, minimum size=1mm, gray]

\begin{tikzpicture}[scale=3]

    \draw (-1,0) arc (180:360:1cm and 0.2cm);
    \draw[dashed, gray] (-1,0) arc (180:0:1cm and 0.2cm);
    
    \draw (0,1) arc (90:270:0.25cm and 1cm);
    \draw[dashed, gray] (0,1) arc (90:-90:0.25cm and 1cm);
    
    \draw (0.50, 0.86) arc  (90:270:0.18cm and 0.86cm);
    \draw[dashed, gray] (0.50, 0.86) arc  (90:-90:0.18cm and 0.86cm);
    
    \draw (0,0) circle (1cm);
    \shade[ball color=blue!10!white,opacity=0.20] (0,0) circle (1cm);
    
    \node (O) at (0, 0) [marknode] {};
    \node (O1) at (-3, 0) [marknode] {};
    \node (O2) at (0.50, 0) [marknode] {};
    \node (N1) at (0.50, 0.86) {};
    
    \node (P) at (0.33, 0.32) [marknode] {};
    \node (P1) at (0.325, -0.19) [marknode] {};
    \node (P2) at (0.375, -0.14) {};
    
    \node (OO) at (0-2, 0-0.7) [marknode] {};
    \node (Oh) at (0-2, 0.3-0.7)  {};
    \node (Ov) at (0.3-2, 0-0.7)  {};
    \node (On) at (0-2-0.2, 0-0.7-0.2) {};
    
    \draw[->, thick] (O1) -- (O) node [midway, below] {$\mathbf v_M$};
    \draw[->, thick] (O) -- (P) node [pos=0.5, left] {$\mathbf v_\infty^+$};
    \draw[->, thick] (O1) -- (P) node [midway, above] {$\mathbf v_{SC}^+$};
    
    \draw[dashed, gray, thin] (O) -- (0.50*1.2, 0.86*1.2);
    \draw[dashed, gray, thin] (O2) -- (P);
    \draw[dashed, gray, thin] (O2) -- (P1);
    \draw[dashed, gray, thin] (O) -- (P1);
    \draw[dashed, gray, thin] (O1) -- (P1);
    
    \draw[->] (1.1, 0) arc  (0:60:1.1) node [midway, right] {$\alpha$};
    \draw[->] (P2) arc  (190:165:0.18cm and 0.86cm) node [midway, right] {$k$};
    
    \draw[->] (OO) -- (Ov) node [right] {$\hat{\boldsymbol i}_v$};
    \draw[->] (OO) -- (Oh) node [above] {$\hat{\boldsymbol i}_h$};;    
    \draw[->] (OO) -- (On) node [left] {$\hat{\boldsymbol i}_n$};;

\end{tikzpicture}

\end{center}
\caption{The pull and crank angles, $\alpha$ and $k$ respectively, fully define the outbound spacecraft velocity $\mathbf v_{SC}$.\label{fig:pullcrank}}
\end{figure}